# Do Students with Different Personality Traits Demonstrate Different Physiological Signals in Video-based Learning?

Chun-Hsiung Tseng[1,2], Hao-Chiang Koong Lin[3], Yung-Hui Chen[4], Jia-Rou Lin[1], and Andrew Chih-Wei Huang[5]

## Abstract

Past researches show that personality trait is a strong predictor for one's academic performance. Today, mature and verified marker systems for assessing personality traits already exist. However, marker systems-based assessing methods have their own limitations. For example, dishonest responses cannot be avoided. In this research, the goal is to develop a method that can overcome the limitations. The proposed method will rely on physiological signals for the assessment. Thirty participants have participated in this experiment. Based on the statistical results, we found that there are correlations between students' personality traits and their physiological signal change when learning via videos. Specifically, we found that participants' degree of extraversion, agreeableness, conscientiousness, and openness to experiences are correlated with the variance of heart rates, the variance of GSR values, and the skewness of voice frequencies, etc.

## 1. Motivation and Background

### 1.1 The Background of Using Personality Traits to Help Students

Why do we need a method to assess students' personality traits? Existing literature has highlighted the relationship between personality traits and learning style or performance. For example, the research of Komarraju et al. pointed out that the Big Five (a popular model for the assessment of personality traits) together explained 14% of the variance in GPA while learning styles explained an additional 3% (Komarraju et al., 2011). Knowing students' personality traits in advance may help teachers assist students when they run into trouble in learning. Furthermore, utilizing personality traits as information can make it easier to adapt to students' preferred learning styles and may achieve better learning performance in turn. Among all learning mechanisms, we would like to limit the scope of this research to video-based learning. The spread of Covid-19, which is perhaps the most influential event in recent years, makes most college students in Taiwan experience video-based learning since mid-2021. In such a scenario, it is more difficult for lecturers to interact with students and hence figuring out a more efficient way to understand students becomes a more urgent need.

Learning via videos is not a completely new concept. Furthermore, since students learning via videos can learn at their own pace to some degree, it brings chances to researchers to understand students' learning behaviors. For example, the research results of Brinton et al. showed that for courses using the MOOC (Massive Open Online Courses) method, students' video-watching behaviors can be a good clue for figuring out when they need help from instructors (Brinton et al., 2016).

[1] Department of Electrical Engineering, YuanZe University, R.O.C.
[2] Chun-Hsiung Tseng is the corresponding author of this manuscript.
[3] Department of Information and Learning Technology, National University of Tainan, R.O.C.
[4] Department of Computer Information and Network Engineering, Lunghwa University of Science and Technology
[5] Department of Psychology, Fo Guang University, R.O.C.

1. Foshee, Elliott, and Atkinson pointed out that adopting an adaptive learning mechanism has significant positive effects on students' learning and academic competence (Foshee, Elliott, & Atkinson, 2016). In most scenarios, an e-learning environment is needed to adopt adaptive learning methods. Kerr defined adaptive learning as "...a way of delivering learning materials online, in which the learner's interaction with previous content determines... (Kerr, 2016)." As pointed out by Brinton et al., students' video-watching behaviors can be a good clue for figuring out when they need help from instructors (Brinton, 2016). The behaviors being analyzed in the research of Brinton et al. were play, pause, skip back, skip forward, rate-change fast, rate-change slow, and rate-change default.

According to Duff et al., learners with different personality traits tend to use different learning strategies (Duff et al., 2004). As a result, if we can assess students' personality traits via data recorded in video-based learning, we will have more information to adapt to students' needs.

## 1.2 Limitations of Existing Methods

Certainly, there were already mechanisms developed for personality traits assessment. A marker system which is a transparent format to quantify one's personality traits in terms of the degree of Extraversion, Agreeableness, Conscientiousness, Emotional stability, and Imagination was developed by Goldberg in 1992 (Goldberg, 1992). However, using questionnaires does have some drawbacks. Perhaps, dishonest responding is one of the most severe problems (Zettler et al., 2015). Besides, filling and interpreting questionnaire results takes additional time. These issues limit the usage of existing methods.

## 1.3 The Contribution and Summarization of the Proposed Method

In this research, the goal is to develop a method to assess personality traits without questionnaires and marker systems. The design has several advantages:
1. it uses only physiological signals and thus does not have strong dependency on e-learning systems
2. there is no need to spend time filling questionnaires and doing analysis
3. there is a chance that physiological signals can reflect psychological status of human beings in a real-time manner
4. knowing students' personality traits helps to adapt to their needs as well as adjusting teaching pedagogies
5. the chance of dishonest responding is reduced

In addition to the statistical results, we also developed a device prototype, a software for collecting the physiological signals, and a mathematics model to present the correlation between physiological signals and personality traits. The rest of this manuscript is summarized as the following:
1. section 2: summary of related works
2. section 3: explain the design of the experiment
3. section 4: describe the analysis process and results of the collected data
4. section 5: conclusion and future work

# 2. Related Works

The goal of this research is to explore the relationship between personality traits and physiological signals and to figure out whether it is possible to assess students' personality traits via physiological signals. In this section, the survey of existing literature of personality traits, the connection between physiological signals and psychological status, and the relationship between personality traits and learning styles are presented.

## 2.1 Personality Traits

The research of Goldberg proposed the Big Five factor structure (Goldberg, 1990). Later, a marker system which is a transparent format to quantify one's personality traits in terms of the degree of Extraversion, Agreeableness, Conscientiousness, Emotional stability, and Imagination was also developed (Goldberg, 1992). The Big Five factors has been adopted in many researches. The research of Barrick and Mount found that the Big Five dimensions are highly related with one's job performance (Barrick & Mount, 1991). Rothmann and Coetzer also pointed out that that Emotional Stability, Extraversion, Openness to Experience and Conscientiousness were related to task performance and creativity (Rothmann & Coetzer, 2003). Butt, Khalid, and Satti conducted a survey in the healthcare sector of Pakistan and found that the personality traits Agreeableness, Conscientiousness and Openness to experience (Imagination) had positive impact on organizational effectiveness while Neuroticism had negative impact. O'Connor and Paunonen found that one's level of Conscientiousness is strongly related with academic success; while Openness to Experience was sometimes positively associated with scholastic achievement, whereas Extraversion was sometimes negatively related to the same criterion (O'Connor & Paunonen, 2007).

## 2.2 Physiological Signal and Psychological Status

Existing research results showed that physiological signals can be used as indicators for some psychological status. The research of Villarejo, Zapirain, and Zorrilla built a stress sensor based on Galvanic Skin Response (GSR) (Villarejo, Zapirain, & Zorrilla, 2012). Lykken used GSR to build a lie detector (Lykken, 1959). The research reported a very high correction rate. Hamadicharef proposed a method to measure one's attention level from her/his electroencephalogram recordings (Hamadicharef et al., 2009). The research reported a classification accuracy rate of 89.4%. The research of Guo et al. used principal component analysis and SVM to classify emotion states based on heart rate variability and reached the accuracy of 71.4% (Guo et al., 2016). The research of Sriramprakash, Prasanna, and Murthy successfully developed a model to assess individual's degree of stress with electrocardiogram and GSR (Sriramprakash, Prasanna, & Murthy, 2017). Egger, Ley, and Hanke conducted a survey and found that physiological signals achieved 79.3% accuracy to assess individuals' emotion state and speech recognition achieved 80.46% for happiness and sadness specifically (Egger, Ley, & Hanke, 2019). Khan et al. proposed a novel deep neural network (DNN) architecture based on the fusion of raw heartbeat and breathing data for classifying and visualizing various emotion states and achieved 71.67% accuracy (Khan et al., 2021). Bastos developed several models for assessing personality traits via physiological signals. The signals used in the research were pupil, ECG, BVP (Blood Volume Pulse), and EDA (GSR) and there were totally 473 features extracted from the signals (Bastos, 2019).

Based on the survey of existing literature, it appears that the connection between physiological signals and psychological status has been established. Moreover, as pointed out by Anglim et al., one's personality traits and psychological status are correlated (Anglim et al., 2020), which indirectly shows that there is a possibility to figure out the relationship between one's personality traits and physiological signal change during some events. The research of Kocjan, Kavčič, & Avsec proposed very similar results (Kocjan, Kavčič, & Avsec, 2021). Besides, do we need medical-grade devices to achieve comparable results? According to Kyriakou et al., wearable devices were capable of detecting stress with 84% accuracy (Kyriakou et al., 2019), which suggests that we do not need medical-grade devices during the experiments.

## 2.3 Relationship between Personality Traits and Learning Style/Performance

Komarraju et al. pointed out that the Big Five together explained 14% of the variance in GPA while learning styles explained an additional 3% (Komarraju et al., 2011). Köseoglu reported very similar results, but the ratio grew to 17% and 5% respectively (Köseoglu, 2016). Both researches showed that personality traits do explain one's academic achievements to a certain degree. Furthermore, Duff et al. found that people with a higher degree of extraversion and imagination tend to use the deep learning approach; people with a higher degree of neuroticism and agreeableness tend to use the surface learning approach; people with a higher degree of extraversion and conscientiousness and a lower degree of neuroticism tend to use the strategic learning approach (Duff et al., 2004). Although using a different personality model, the research results of Moreira et al. showed almost identical results. They found that students with a steady temperament or high character coherence tend to use the deep learning approach (Moreira et al., 2021). Moreira, Pedras, and Pombo conducted a survey and found that personality is the strongest predictor of academic performance (Moreira, Pedras, & Pombo, 2020).

# 3. Experiment Design

The goal of this research is to figure out the relationship between personality traits and changes in physiological signals in video-based learning. The hypothesis below was proposed:

*There is a relationship between participants' personality traits and changes in physiological signals in video-based learning.*

If the relationship exists, we can use physiological signals to assess participants' personality traits. In the sections below, the experiment process to verify the hypothesis will be presented.

## 3.1 Collection of Personality Traits

In this research, the IPIP big five factor markers questionnaire[6] was adopted for the analysis of participants' personality traits as the ground truth. The questionnaire has 50 questions and each question has five possible answers: very inaccurate, moderately inaccurate, neither accurate nor inaccurate, moderately accurate, or very accurate. The questionnaire evaluates the scores of the five personality traits: extraversion, agreeableness, conscientiousness, emotional stability, and openness to experiences (imagination). The higher the score a participant received in a personality trait, the stronger the tendency the participant demonstrated in the personality trait. For convenience, we created a Web-based questionnaire application[7] to facilitate participants.

---

[6] https://ipip.ori.org/new_ipip-50-item-scale.htm
[7] https://lendle-ipip.glitch.me/

## 3.2 Collection of Physiological Signals

In this research, participants' voice frequencies and the physiological signals including hear rate values and GSR values were collected as features. We have built a device based on a Raspberry Pi for collecting the needed data. The figure below demonstrates the prototype device:

Figure 1 the device for collecting signals

## 3.3 Experiment Process

The steps below were included in the experiment:

1. participants were randomly partitioned into two groups
2. participants watched a video; the first group of participants watched a video that introduces fundamental concepts of JavaScript while the second group of participants watched a cartoon
3. after the end of the video, participants were asked to summarize her/his understandings of the concepts included in the video
4. participants were asked to fill in the IPIP big five factor markers questionnaire

# 4. Data Analysis

## 4.1 Description of Variables

To capture the personality traits of participants, the IPIP big five factor markers questionnaire was adopted. Each item was then calculated according to the scoring instructions mentioned in the IPIP website, and the scoring results are numbers indicating the strength of each dimension. On the other hand, participants' voice frequencies, heart rates, and GSR values were also captured for correlation analysis. To simplify and analytical process, we did not use these values directly but used the variance, standard deviation, and skewness of them instead.

To sum up, there were two sets of variables in this research:

1. 9 independent variables:
    a) the variance of heart rates (hr_var)
    b) the standard deviation of heart rates (hr_sd)
    c) the skewness of heart rates (hr_sk)
    d) the variance of GSR values (gsr_var)
    e) the standard deviation of GSR values (gsr_sd)
    f) the skewness of GSR values (gsr_sk)
    g) the variance of voice frequencies (fr_var)
    h) the standard deviation of voice frequencies (fr_sd)
    i) the skewness of voice frequencies (fr_sk)
2. 5 dependent variables:
    a) extraversion
    b) agreeableness
    c) conscientiousness
    d) emotional stability
    e) openness to experience

## 4.2 Statistics Results of Personality Traits and Physiological Signals

First, we followed the steps indicated by Goldberg to process the collected questionnaires (Goldberg, 1992). The questionnaire has 50 items and was designed to evaluate the tendency of five personality traits. The tendency of a personality trait is stronger if the calculated score is higher. Furthermore, several methods were adopted to collect voice frequency data, GSR values, and heart rates. Specifically, the *javax.sound.sampled* library was used for collecting the voice frequency data, the Grove GSR sensor[8] was adopted for collecting the GSR values, and the Grove Ear-clip heart rate sensor was utilized for collecting the heart rate values[9]. The list below shows the sampling rate of each signal:

1. voice frequency: 2 samples per second
2. GSR: 0.3 samples per second
3. heart rates: 1.1 samples per second

Then, we grouped each signal by a 30 seconds time interval. For each group, the mean value of data contained in the group was calculated and used as the representative value. After the preprocessing step, each participant had roughly 40 records for each type of signal, since the experiment lasted for roughly 20 minutes. The only exception was the voice frequency data. The voice frequency data was only collected in the last 2 minutes of the experiment, so each participant has only 4 records of the corresponding grouped voice frequency data. The table below shows the statistics results:

Table 1 Statistics results of physiological signals

| Data Category | Statistics | | | |
|---|---|---|---|---|
| | Min | Median | Mean | Max |
| heart rate | 42.88 | 56.69 | 57.39 | 77.28 |
| GSR | 88.83 | 406.1 | 389.40 | 578.40 |
| voice frequency | 0.16 | 5.83 | 5.72 | 16.98 |

## 4.3 Data Analysis

The pearson correlation and the canonical correlation were adopted to figure out the correlation between physiological signals and personality traits. The results were shown in the table below:

Table 2 Correlation Analysis

| | Extraversion | Agreeableness | Conscientiousness | Emotional kekeStability | Openness to Experience |
|---|---|---|---|---|---|
| hr_var | 0.41 | 0.24 | 0.15 | 0.32 | 0.31 |
| hr_sd | 0.42 | 0.20 | 0.13 | 0.31 | 0.29 |
| hr_sk | -0.03 | 0.19 | 0.35 | 0.22 | -0.14 |
| gsr_var | 0.50 | 0.44 | 0.21 | 0.35 | 0.37 |
| gsr_sd | 0.46 | 0.48 | 0.31 | 0.34 | 0.33 |
| gsr_sk | -0.03 | 0.05 | 0.02 | 0.09 | -0.21 |
| hr_var | -0.04 | 0.00 | -0.08 | 0.16 | -0.05 |
| hr_sd | -0.02 | 0.00 | -0.06 | 0.18 | -0.09 |
| hr_sk | -0.13 | 0.17 | 0.39 | 0.09 | 0.16 |

[8] https://wiki.seeedstudio.com/Grove-GSR_Sensor/

[9] https://wiki.seeedstudio.com/Grove-Ear-clip_Heart_Rate_Sensor/

As shown above, medium-level correlation was observed between some pairs of parameters. For example, the correlation coefficient between gsr_var and Extraversion was 0.5, which demonstrates medium-level correlation. The *rcorr* function defined in the *Hmisc* package was then utilized to further verify the significance of the correlation coefficients. The table below shows significantly correlated pairs:

Table 3 The significance of correlations.

| Pearson Correlation | Extraversion | Agreeableness | Conscientiousness | Openness.to.Experience |
|---|---|---|---|---|
| hr_var | 0.4108 * | - | - | - |
| hr_sd | 0.4178 * | - | - | - |
| gsr_var | 0.5034 ** | 0.4389* | - | 0.3659* |
| gsr_sd | 0.4634 ** | 0.4774 ** | - | - |
| fr_sk | - | - | 0.3909* | - |

**: Correlation is significant at the 0.01 level

*: Correlation is significant at the 0.05 level

The following significant correlations have been observed:

1. The correlation between the personality factor "Extraversion" and the variance of heart rates is positive and significant; the strength is moderate.
2. The correlation between the personality factor "Extraversion" and the standard deviation of heart rates is positive and significant; the strength is moderate.
3. The correlation between the personality factor "Extraversion" and the variance of GSR is positive and significant; the strength is moderate.
4. The correlation between the personality factor "Extraversion" and the standard deviation of GSR is positive and significant; the strength is moderate.
5. The correlation between the personality factor "Agreeableness" and the variance of GSR is positive and significant; the strength is moderate.
6. The correlation between the personality factor "Agreeableness" and the standard deviation of GSR is positive and significant; the strength is moderate.
7. The correlation between the personality factor "Conscientiousness" and the skewness of voice frequencies is positive and significant; the strength is weak.
8. The correlation between the personality factor "Openness to Experience" and the variance of GSR is positive and significant; the strength is weak.

Four of five personality factors were significantly correlated with at least one physiological signal. Besides, the results show that the variance of GSR alone is significantly correlated with "Extraversion", "Agreeableness", and "Openness.to.Experience".

However, "Emotional Stability" was not significantly correlated with any physiological signal. Additionally, the canonical correlation analysis was also performed. Since there were 9 physiological signals and 5 personality traits, at most five canonical dimensions could exist. Wilks' Lambda test showed dimensions 1-5, 2-5, 3-5, 4-5, and 5-5 were all not significant. The results are shown in the table below:

Table 4 Wilks' Lambda Test

| Dimension | Canonical Correlation | df1 | df2 | p |
|---|---|---|---|---|

| 1-5 | 0.129 | 45 | 74.67 | 0.5468969 |
| 2-5 | 0.375 | 32 | 64.29 | 0.9342119 |
| 3-5 | 0.691 | 21 | 52.24 | 0.9955165 |
| 4-5 | 0.879 | 12 | 38 | 0.9969680 |
| 5-5 | 0.957 | 5 | 20 | 0.9674491 |

However, Roy's Test showed that the first dimension was significant:

Table 5 Roy's Test

| Dimension | Canonical Correlation | df1 | df2 | p |
|---|---|---|---|---|
| 1 | 0.655 | 5 | 24 | 5.739819e-05 |

The table below shows the standardized coefficients of the first dimension:

Table 6 The Standardized Coefficients of the First Dimension

| | Standardized Coefficient |
|---|---|
| Physiological Signal | |
| hr_var | 0.0921 |
| hr_sd | -0.4845 |
| hr_sk | 0.0119 |
| gsr_var | 0.2601 |
| gsr_sd | -1.1349 |
| gsr_sk | 0.1907 |
| fr_var | -0.7488 |
| fr_sd | 0.7137 |
| fr_sk | -0.3755 |
| Personality Traits | |
| Extraversion | -0.3894 |
| Agreeableness | 0.0140 |
| Conscientiousness | -0.3907 |
| Emotional.Stability | -0.4915 |
| Openness.to.Experience | -0.4501 |

To assess the importance of features, the distribution of the standardized coefficients was then analyzed. The absolute value of coefficients reflects the strength of the feature, so the quantile position of the absolute coefficients was used. For the x axis, the 75th percentile was 1.0440 and for the y axis, the 75th percentile was 0.8489. The absolute value of the standardized coefficients of gsr_sd, fr_var, fr_sd, Emotional.Stability, and Openness.to.Experience are larger than the third quartile, which draws to the conclusion that gsr_sd, fr_var, and fr_sd are significantly correlated with "Emotional Stability" and "Openness to Experience". The result shows that when combining with other personality traits, the relationship between "Emotional Stability" and physiological signals is still observable.

Based on the analysis results, we can say that the hypothesis proposed in section 3 holds, i.e., participants' physiological signals reveal their personality traits. Besides, the functionality of the device we designed is self-contained, which means adopting the proposed method and device in an existing lecturing environment will help lecturers figuring out students' personality traits without interfering with other technologies current deployed.

## 4.4 Effects of Different Types of Materials

Since there were two groups of participants watching different types of videos, it appears there is a need to investigate the impacts of types of videos. T-test was performed to evaluate the difference of physiological signals between the two groups of participants. The results are shown below:

Table 7 T-test Results between the Two Groups

| | | levene test | estimate | t | p | df | confidence | |
|---|---|---|---|---|---|---|---|---|
| | | | | | | | high | low |
| hr_var | equal variance | 0.023 | -20.871 | -1.569 | 0.128 | 28.000 | 6.376 | -48.117 |
| | unequal variance | | -20.871 | -1.569 | 0.129 | 25.183 | 6.514 | -48.255 |
| hr_sd | equal variance | 0.097 | -1.339 | -1.285 | 0.209 | 28.000 | 0.795 | -3.473 |
| | unequal variance | | -1.339 | -1.285 | 0.210 | 26.366 | 0.801 | -3.479 |
| hr_sk | equal variance | 0.640 | 0.060 | 0.138 | 0.891 | 28.000 | 0.947 | -0.827 |
| | unequal variance | | 0.060 | 0.138 | 0.891 | 26.948 | 0.948 | -0.829 |
| gsr_var | equal variance | 0.001 | -4,141.757 | -1.719 | 0.097 | 28.000 | 794.547 | -9,078.06 |
| | unequal variance | | -4,141.757 | -1.719 | 0.106 | 14.822 | 1,000.053 | -9,283.56 |
| gsr_sd | equal variance | 0.019 | -21.178 | -1.455 | 0.157 | 28.000 | 8.627 | -50.984 |
| | unequal variance | | -21.178 | -1.455 | 0.163 | 17.810 | 9.415 | -51.771 |
| gsr_sk | equal variance | 0.029 | -0.706 | -2.233 | 0.034 | 28.000 | -0.058 | -1.353 |
| | unequal variance | | -0.706 | -2.233 | 0.036 | 22.794 | -0.052 | -1.360 |
| fr_var | equal variance | 0.069 | -4.011 | -1.585 | 0.124 | 28.000 | 1.173 | -9.196 |
| | unequal variance | | -4.011 | -1.585 | 0.127 | 21.794 | 1.240 | -9.263 |
| fr_sd | equal variance | 0.153 | -0.944 | -1.836 | 0.077 | 28.000 | 0.109 | -1.996 |
| | unequal variance | | -0.944 | -1.836 | 0.078 | 25.485 | 0.114 | -2.001 |
| fr_sk | equal variance | 0.181 | 0.110 | 0.671 | 0.508 | 28.000 | 0.446 | -0.226 |
| | unequal variance | | 0.110 | 0.671 | 0.508 | 26.812 | 0.447 | -0.227 |

According to the results, the difference of gsr_sk between the two groups is significant, while the differences of other signals are not significant. As a result, it appears reasonable to conclude that the result of this research is not affected by the different types of video materials.

## 4.5 Limitations of the Results

The current results have several limitations:

1. The current results focus on figuring out the correlations between personality traits and physiological signals, not prediction models.
2. In the experiment, participants watched a 20-minutes video, so the results may not be directly applicable in ordinary classrooms.

# 5. Conclusion and Future Work

The major contribution of this research is that we found that participants' physiological signal change during a video-based learning session reflects their personality traits. Specifically, we discovered the following (positive) correlations:

Table 8 Correlations between Personality Traits and Physiological Signals

|  | Variance of Heart Rates | Variance of GSR Values | Skewness of Voice Frequencies |
|---|---|---|---|
| Extraversion | √ | √ | |
| Agreeableness | | √ | |
| Conscientiousness | | | √ |
| Openness to Experience | | √ | |

Past researches have already shown that there are connections between one's personality traits and psychological status. According to Sriramprakash, Prasanna, and Murthy, individual's level of stress can be detected with electrocardiogram and GSR (Sriramprakash, Prasanna, & Murthy, 2017). Guo et al. reported that using heart rate variability alone can classify participants' emotion states (Guo et al, 2016). Crifaci et al. pointed out the variance of GSR between the stressed and unstressed groups was significantly different (Crifaci et al., 2013). Furthermore, Vollrath's research proved that people's personality affects how they react to stress (Vollrath, 2001). Compared with existing research results, our findings appear to be reasonably consistent. Moreover, compared with traditional personality traits assessment methods, which usually rely on questionnaires, the proposed method can be more objective and efficient since there is less human intervention.

Our results are partially consistent with Bastos' results (Bastos, 2019). Bastos found that one's degree of extraversion can be assessed by her/his ECG and EDA. ECG is strongly related to heart rates variance[10], while EDA is identical with GSR[11]. The two research results appear to be similar in assessing one's degree of extraversion. Additionally, both research results showed that one's degree of openness to experience can be assessed by variance in GSR values. Besides, our method used a much smaller set of features (9 v.s. 145 for assessing extraversion).

In the future, the goal is to develop an assessment tool/device that can be used in classroom based on the current result. Knowing students' personality traits brings opportunities to improve their learning performance. The research of Komarraju et al. (Komarraju et al., 2011) and Köseoglu (Köseoglu, 2016) showed the connections between students' personality traits and their academic achievements. With such a tool, teachers may have a clue about the possible weaknesses in students' learning style and then support them before they fail. Furthermore, as pointed out by Gadekallu et al., the metaverse expands the scope of capabilities in social media (Gadekallu et al., 2022). As a result, with the security convern being carefully addressed, how users behave and interact in the metaverse can become another potential source of features for investigating their personality traits.

---

[10] https://www.ncbi.nlm.nih.gov/pmc/articles/PMC7472094/

[11] https://en.wikipedia.org/wiki/Electrodermal_activity

## Acknowledgment

This research is partially supported by the "Judging personality traits by physiological signals and constructing a digital learning environment that adapts to individual personality traits: the impact on learning effectiveness, achievement emotions, and student engagement." project, which was funded by the Ministry of Science and Technology, Taiwan, R.O.C. under Grant no. MOST 110-2511-H-155-004.